\documentclass[runningheads]{llncs}

\usepackage[misc]{ifsym}
\usepackage{graphicx}
\usepackage{amsmath}
\usepackage{subfig}
\usepackage{booktabs} 
\usepackage{diagbox}
\usepackage{threeparttable}
\usepackage{multirow}

\usepackage{stackengine}
\newcommand\xrowht[2][0]{\addstackgap[.5\dimexpr#2\relax]{\vphantom{#1}}}
\usepackage{url}

\begin{document}
\title{Domain Knowledge Based Brain Tumor Segmentation and Overall Survival Prediction}
\titlerunning{Brain Tumor Segmentation and OS Prediction}
%
\author{Xiaoqing Guo\inst{1} \and
Chen Yang\inst{1} \and
Pak Lun Lam\inst{2} \and
Peter Y.M. Woo\inst{3} \and
Yixuan Yuan\inst{1}\textsuperscript{(\Letter)}}
\authorrunning{X. Guo et al.}
%
\institute{Department of Electrical Engineering, City University of Hong Kong, Hong Kong SAR, China \and Department of Diagnostic and Interventional Radiology, Kwong Wah Hospital, Hong Kong SAR, China \and Department of Neurosurgery, Kwong Wah Hospital, Hong Kong SAR, China
\email{yxyuan.ee@cityu.edu.hk}}

\maketitle              
\begin{abstract}
Automatically segmenting sub-regions of gliomas (necrosis, edema and enhancing tumor) and accurately predicting overall survival (OS) time from multimodal MRI sequences have important clinical significance in diagnosis, prognosis and treatment of gliomas. However, due to the high degree variations of heterogeneous appearance and individual physical state, the segmentation of sub-regions and OS prediction are very challenging. To deal with these challenges, we utilize a 3D dilated multi-fiber network (DMFNet) with weighted dice loss for brain tumor segmentation, which incorporates prior volume statistic knowledge and obtains a balance between small and large objects in MRI scans. For OS prediction, we propose a DenseNet based 3D neural network with position encoding convolutional layer (PECL) to extract meaningful features from T1 contrast MRI, T2 MRI and previously segmented sub-regions. Both labeled data and unlabeled data are utilized to prevent over-fitting for semi-supervised learning. Those learned deep features along with handcrafted features (such as ages, volume of tumor) and position encoding segmentation features are fed to a Gradient Boosting Decision Tree (GBDT) to predict a specific OS day. 

\end{abstract}
\section{Introduction}

The annual incidence of primary brain tumors is increasing and poses a significant burden on public health \cite{bray2018global}. Glial cells comprise approximately half of the total volume of the brain with a glial cell-to-neuron ratio of 1:1 \cite{von2016search}. They are principally responsible for maintaining homeostasis, providing support and protecting neurons. Gliomas are the most common primary brain tumors in humans and originate from glial cells, accounting for 35\% to 60\% of all intracranial tumors \cite{bray2018global}. The age-standardized incidence rate of gliomas is 4.7 per 100000 person-years and in clinical practice the diagnosis of such tumors requires neurosurgery to obtain a tissue biopsy, which entails considerable risks for patients. Moreover, according to the World Health Organisation (WHO), gliomas can be histologically classified to into four grades, with each resulting in distinctly different durations of overall survival (OS). Although high-grade gliomas (HGG), i.e. WHO grade III or IV, are considered more aggressive, there is a growing body of evidence that such a histopathological classification is inadequate to prognosticate OS due to the nuanced variations in molecular profile from one tumor to the next \cite{van2010interobserver}. By incorporating the glioma pathological diagnosis data, segmenting tumor sub-regions exhibited by magnetic resonance imaging (MRI) has been known to provide additional quantitative information for OS prediction. However, the process of manual image segmentation is highly time-consuming, often requires experienced neuro-radiologists and can be subject to inter-observer variations. To address these issues, developing an automated accurate segmentation tool that can reliably detect OS-relevant imaging biomarkers is urgently needed. 



Gliomas, especially HGGs, often possess intratumoral heterogeneity that could represent different MRI signal intensity profiles across multi-modality imaging sequences \cite{menze2014multimodal}. Over the last decade, a number of scholars have proposed algorithms to automatically segment these glioma sub-regions in order to determine an accurate preoperative prognosis with varying degrees of success \cite{feng2018brain,kao2018brain,nie2019multi,wang2015radiological,zhao2018automated}. Promising progress has been made using traditional machine learning methods  \cite{wang2015radiological,zhao2018automated}, which calculated low-level handcrafted,  radiological features to describe images and trained a classifier or a regression for tumor segmentation and OS prediction. These handcrafted features were usually defined by experienced neuro-radiologists founded on prior knowledge of the exact histological diagnosis of the glioma and could have been a potential source of bias. This simple, straight-forward feature analysis approach potentially also disregarded a great deal of useful information embedded within the MR images, prohibiting the full effective utilization of sub-region segmentation for OS prediction. Recently, deep learning methods have demonstrated superior image processing capabilities that have been proven to effectively overcome these limitations \cite{feng2018brain,kao2018brain,lecun2015deep,nie2019multi}. Instead of defining handcrafted features, Convolutional Neural Network (CNN) methods jointly trains feature extractor and classifier to adaptively derive high-yield information and enhance model performance \cite{lecun2015deep}. Inspired by the superior outcomes of this methodology, researchers are increasingly applying CNN for brain tumor segmentation and patient OS prediction \cite{feng2018brain,kao2018brain,nie2019multi}. Nie et al. \cite{nie2019multi} adopted a VGG-based network to automatically extract high-yield features from gadolinium contrast-enhanced T1 (T1ce) and diffusion tensor imaging sequences, and then utilized the extracted features together with tumor volume data to train a support vector machine for final OS prediction. Kao et al. \cite{kao2018brain} incorporated location information from a brain parcellation atlas to obtain accurate glioma segmentation results. Kao et al. \cite{kao2018brain} also analyzed connectome tractography information to identify potentially tumor-induced damaged brain regions and demonstrated that incorporation of this feature dataset resulted in superior OS prediction than including customary age, volumetric, spatial and morphological features alone. 

Despite the relatively good performance of automatic tumor segmentation, the results of OS prediction are far from satisfactory  \cite{bakas2018identifying,menze2014multimodal}. For example, the patient OS prediction model of the first-ranking team \cite{feng2018brain} of the Brain Tumor Segmentation (BraTS) 2018 challenge, an international competition with open-source MRI Digital Images in Communications in Medicine (DICOM) data organized by the School of Medicine of the University of Pennsylvania, only resulted in a accuracy of 0.62  \cite{bakas2018identifying}. Two factors may have resulted in the limited predictive capacity of previous deep learning methodologies. First, \cite{kao2018brain} was the first effort to incorporate tumor location-based information in the CNN for brain tumor segmentation. However, such location data, which is crucial for OS prediction, has not been considered to predict OS. 
Secondly, pre-existing algorithms for OS prediction were usually based on supervised CNN models. However, the limited available number of datasets, led to considerable over-fitting problems \cite{feng2018brain,kao2018brain,nie2019multi}. In contrast, unlabeled MRI DICOM data are readily accessible in the clinical setting. Therefore, making adequate use of such data during the training process could be a promising strategy to improve OS prediction. 

In this paper, we present a 3D dilated multi-fiber network (DMFNet) trained with weighted dice loss to segment glioma sub-regions from MRI scans. Then these predicted segmentation results are combined with T1 contrast and T2 MRI together as inputs for the proposed PECL-DenseNet to extract high-level and meaningful features that is trained with unlabeled as well as labeled data. In addition, we combine the extracted deep features from PECL- DenseNet with handcrafted features (age, tumor volume, volume ratio, surface area, surface area to volume ratio, location of the tumor's epicenter, its corresponding brain parcellation, relevant location of the tumor epicenter to the brain epicenter and resection status) and position encoding segmentation features to train the Gradient Boosted Decision Tree (GBDT) regression for patient OS prediction.

\section{Methodology}
\subsection{Dataset}
The Brain Tumor Segmentation (BraTS) 2019 dataset \cite{bakas2017segmentationGBM,bakas2017segmentationLGG,bakas2017advancing,bakas2018identifying,menze2014multimodal} provides 335 training subjects, 125 validation subjects and 167 testing ones, each with four MRI modality sequences (T1, T1ce, T2 and FLAIR). All the training data have corresponding pixel-level annotations, including necrosis and non-enhancing tumor, edema, and enhancing tumor sub-regions. Partial training data have corresponding subject-level annotations, indicating the OS duration and the resection status, respectively. In particular, only HGG patients with gross total resection (GTR) were evaluated, since resection status is the only consistent modifiable treatment predictor for OS, and only 101 (30\%) training subjects are eligible. There remaining 109 brain tumor training subjects do not undergo GTR have OS data and 49 subjects miss OS labels. 29 validation subjects and 107 testing ones have complete subject-level annotations.

\subsection{Brain Tumor Segmentation}
\begin{figure}[tp]
\centering
\includegraphics[width=123mm]{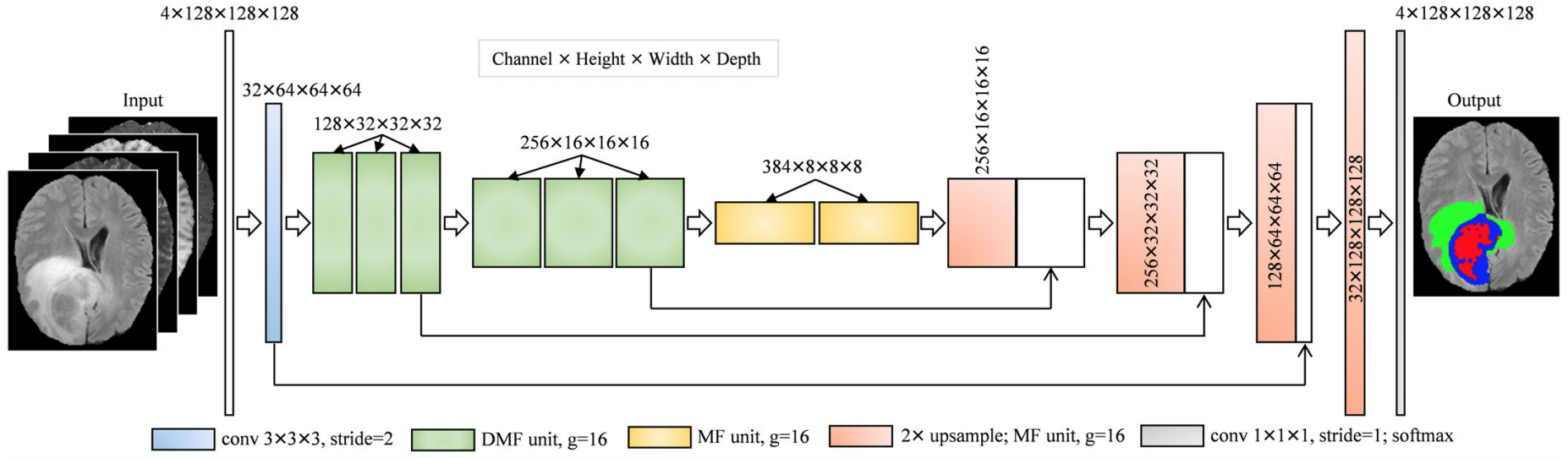}
\caption{Illustration of DMFNet framework for brain tumor segmentation.}
\label{DenseNet}
\end{figure}

A major difficulty with the existing BraTS segmentation challenge is the high computational cost required, since each subject has four modality MRI scans. To tackle this dilemma, our segmentation model is primarily based on DMFNet \cite{chen20193d}, which can significantly reduce the computational cost of 3D networks by an order of magnitude. It slices a complex neural network into an ensemble of lightweight networks or fibers, and further incorporates multiplexer modules to facilitate information flow between fibers. To enlarge the respective field and to capture the multi-scale 3D spatial correlations, DMFNet adds dilated convolution to multiplexer modules.

The accuracy of contrast-enhancing tumor segmentation is usually the worst, compared with peri-tumoral edema and intratumoral necrosis regions, since tumor tissue enhancement often constitutes the smallest volume of the entire tumor. Therefore, we introduce prior volume knowledge to traditional dice loss to resolve this imbalanced class problem, namely by weighted dice loss. In particular, we apply the reciprocal of each tumor volume as our dice weight, given as 0.38, 0.15, 0.47 for necrosis, edema and enhancing tumor, respectively.

\subsection{Overall Survival Prediction}
Features extracted from deep CNN, handcrafted features and position encoding segmentation features are incorporated for the OS prediction. The geometry and location of tumor are crucial for the OS prediction \cite{perez2017glioblastoma}. Therefore, we propose a PECL-DenseNet with considering the location information to extract meaningful features and make adequate use of unlabeled data to prevent over-fitting. With the calculated segmentation result, we define 36 handcrafted features that involves geometry and location information for accurate OS prediction. Moreover, we apply the pooling operator to predicted segmentation to accurately obtain the tumor location information. GBDT regression is performed to fit with normalized features of 210 training data that is with OS labeling. The source code for extracting the handcrafted features and implementing GBDT regression is available at \url{https://github.com/Guo-Xiaoqing/BraTS_OS}.

\subsubsection{PECL-DenseNet}

From T1ce, T2 sequence MRI images and the predicted sub- region segmentation from DMFNet, deep features are extracted by alternate-cascaded 3D dense blocks and transition layers. The dense connectivities in dense blocks can combine information from different convolutional layers, therefore encourage feature reuse and ensure maximum information flow between layers. Specifically, our proposed framework includes four dense blocks as shown in Fig. \ref{DenseNet}. Each block is comprised of 7 densely connected layers, and every layer consists of a batch normalization, a ReLU, and the proposed PECL module. Then the deep features are concatenated with the resection status to derive a five-class OS prediction classification. Specifically, we utilize a digit to represent resection status, given as GTR (2), STR (1) and NA (0). The dimension of extracted image features are reduced to 50 by principal components analysis for further processing.

\begin{figure}[tp]
\centering
\includegraphics[width=123mm]{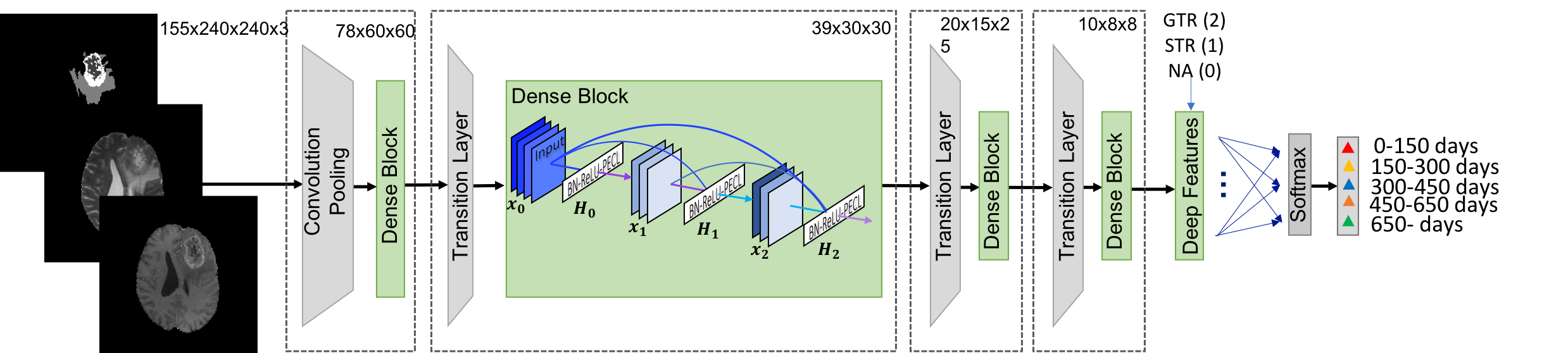}
\caption{Illustration of the proposed PECL-DenseNet for OS prediction. T1 contrast, T2 MRI images and the predicted sub-regions segmentation from DMFNet are concatenated as the input of the PECL-DenseNet. Deep features extracted from the PECL-DenseNet are then combined with resection status to make a five-classes prediction.}
\label{DenseNet}
\end{figure}

\begin{figure}[t]
\centering
\includegraphics[width=123mm]{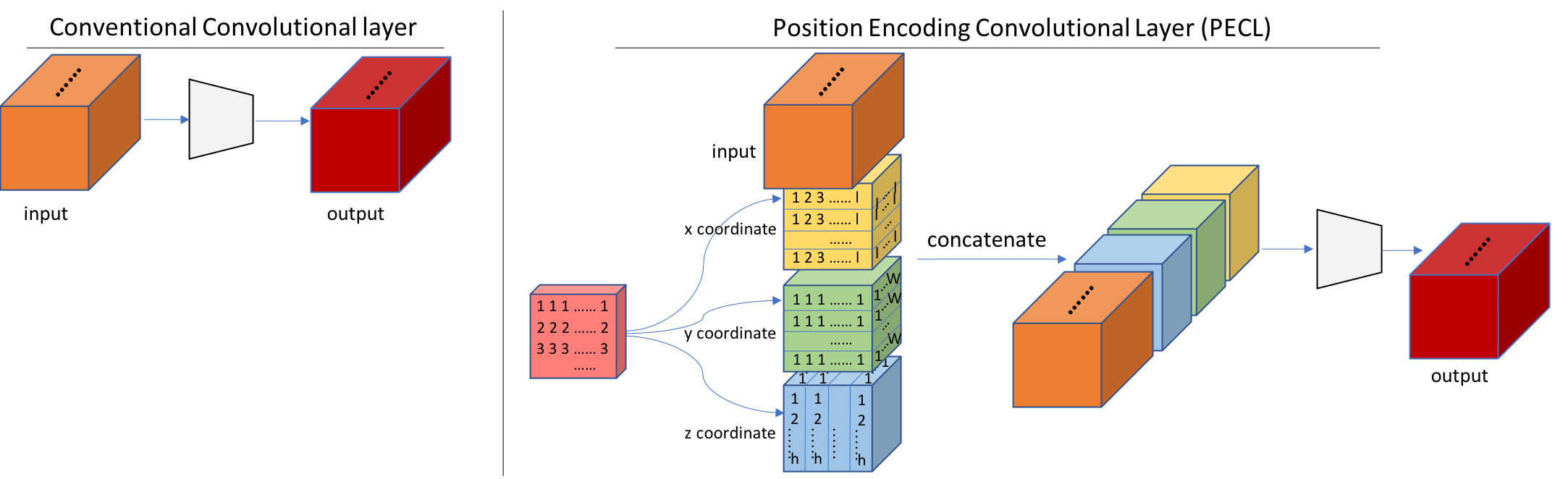}
\caption{Comparison of 3D convolutional layer and the proposed PECL.}
\label{PECL}
\end{figure}

We propose a position encoding convolutional layer (PECL) to incorporate location knowledge for OS prediction. Due to the translation invariance of convolution and global average pooling operator, the extracted deep feature vector usually ignores the spatial information. However, the tumor location is essential for the diagnosis and prognosis of gliomas, especially for HGG. In this regard, we extend the conventional 3D convolutional layer to PECL by incorporating the position information as in Fig. \ref{PECL}. In contrast from the standard convolutional layer, PECL introduces three extra channels ($x, y, z$) to derive a 3D Cartesian coordinate. The introduced channels are individually normalized by dividing their maximum value. Then input feature maps are concatenated with these additional channels for further processing.

To make adequate use of the limited available labeled data and additional unlabeled data in our hand, we develop an effective loss function for semi-supervised learning. Assuming the training set is $\mathcal{D}$ consisting of $N$ samples. Denoting $\mathcal{L} = \left\{(x_{i},y_{i}) \right\} _{i=1}^{L}$ is labeled dataset and $\mathcal{U} = \left\{x_{i} \right\} _{i=L+1}^{N}$ is unlabeled dataset. We aim to learn an OS prediction network parameterized by $\Theta$ through optimizing the following loss function: 

\begin{equation}
\begin{aligned}
\begin{split}
\mathit{L} =  \frac{1}{N}\sum_{i=1}^{N} (\alpha \cdot \mathop {\frac{(n-1)\left \| z_{i}-\mathbf{c}_{y_{i}} \right \|}{\sum_{j\neq y_{i}}^{n}\left \| z_{i}-\mathbf{c}_{j} \right \|}  }\limits_{x\in \mathcal{L}} - \beta \cdot \mathop {y_{i} \log  p_{i}}\limits_{x\in \mathcal{L}} - \gamma \cdot \mathop { p_{i} \log  p_{i}}\limits_{x\in \mathcal{U}}), 
\label{network_loss}
\end{split}
\end{aligned}
\end{equation}
where $p_{i} = \frac{e^{W^{\top}z_{i}+b}}{\sum_{j=1}^{n}e^{W^{\top}_{j}z_{i}+b_{j}}}$. $W_{j}$ is the weight for $j^{th}$ class in fully connected layer and $b_{j}$  is bias. $z_{i}$ denotes extracted features of $i^{th}$ samples and $\mathbf{c}_{j}$ is the $j^{th}$ class feature centroid. N and n represent batch size and number of classes. $\alpha, \beta, \gamma$ are set as 0.5, 1 and 0.1, respectively. The first term is inspired by \cite{wen2016discriminative} and aimed to enforce the extracted features to approximate their corresponding feature centroid and to distance away from other centroids. The accumulative feature centroids are updated by formulation: $\mathbf{c}_{j}^{t+1} = \mathbf{c}_{j}^{t} - 0.5 \cdot \frac{\sum_{i=1}^{N}\delta(j=y_{i})\cdot (\mathbf{c}_{j}-x_{j})}{1+\sum_{i=1}^{N}\delta (j=y_{i})}$, where $t$ denotes sequential iterations. $\delta({\bf \cdot}) = 1$ if condition is satisfied, and $\delta({\bf \cdot}) = 0$ if not. The second term is softmax cross entropy loss for labeled data, and the third one is an information entropy loss for unlabeled data.

\subsubsection{handcrafted feature}

We define 36 handcrafted features that involves non-image features and image features. Non-image features includes age and resection status. In particularly, a two dimensional feature vector is used to represent resection status, given as GTR (1, 0), STR (0, 1) and NA (0, 0). With the calculated segmentation from DMFNet, we calculate 34 image features including volume ($V_{whole}$, $V_{necrosis}$, $V_{edema}$, $V_{enhancing}$), volume ratio ( $\frac{V_{whole}}{V_{brain}}$,  $\frac{V_{necrosis}}{V_{brain}}$,  $\frac{V_{edema}}{V_{brain}}$, $\frac{V_{enhancing}}{V_{brain}}$, $\frac{V_{necrosis}}{V_{enhancing}}$,  $\frac{V_{edema}}{V_{enhancing}}$,  $\frac{V_{necrosis}}{V_{edema}}$), surface area ($S_{whole}$, $S_{necrosis}$, $S_{edema}$, $S_{enhancing}$), surface area to volume ratio ( $\frac{S_{whole}}{V_{whole}}$, $\frac{S_{necrosis}}{V_{necrosis}}$, $\frac{S_{edema}}{V_{edema}}$, $\frac{S_{enhancing}}{V_{enhancing}}$), position of the whole tumor epicenter  (3 coordinates and its corresponding brain parcellation), position of the enhancing tumor epicenter  (3 coordinates and its corresponding brain parcellation), relevant location of the whole tumor epicenter  to brain epicenter  (3 coordinates) and relevant location of the enhancing tumor epicenter  to brain epicenter  (3 coordinates). Note that $V$ and $S$ indicate volume and surface area. $whole$, $necrosis$, $edema$, $enhancing$ and $brain$ denote the entire tumor, necrosis and non-enhancing tumor, edema, enhancing tumor and the entire brain, respectively. To obtain the brain parcellation for tumor epicenter location, we register all the data to LPBA40 atlas \cite{shattuck2008construction}, and 56 different brain parcellations are delineated. 

\subsubsection{Position encoding segmentation}
To reserve the tumor location information, we apply a pooling operator to the predicted segmentation, where the kernel of the pooling operator is $5\times 12\times 12$ and the resolution of predicted segmentation is $155\times 240\times 240$. Thus, a 12400-dimensional feature vector is obtained.

\section{Experiments and Results}

\subsection{Experiment Setup}
In the brain tumor segmentation experiment, we trained the DMFNet with all the 335 training subjects and evaluated on 125 validation subject data and 167 testing data. We used a batch size of six and trained the DMFNet on two parallel Nvidia GeForce 1080Ti GPUs for 300 epochs. The initial learning rate was set as 0.001. During the training phase, we randomly cropped the data into $128\times128\times128$ for training data augmentation. In the testing phase, we utilized zero padding to make the resolution of input MRI data $240\times240\times160$. 

In the OS prediction experiment, we made subjects with OS labels ($101+109=210$ subjects) as labeled data, and regarded the remaining 49 subjects without OS labels and 96 validation subjects as unlabeled data for the 3D PECL-DenseNet training in a semi-supervised strategy. Results of OS prediction were evaluated on 29 validation subjects and 107 testing ones. During 3D PECL-DenseNet training, the initialized learning rate was set to 0.1, and was dropped by 0.1 at 150 and 250 epochs, respectively. All training steps for labeled data and unlabeled data both use batch size of four. Both handcrafted features and location encoding segmentation features are extracted from 210 training subjects, 29 validation subjects and 107 testing ones, which are then combined with deep features extracted from PECL-DenseNet to feed into the GBDT for training and testing, respectively.

\subsection{Results}

\begin{figure}[ht!]
\centering
\includegraphics[width=125mm]{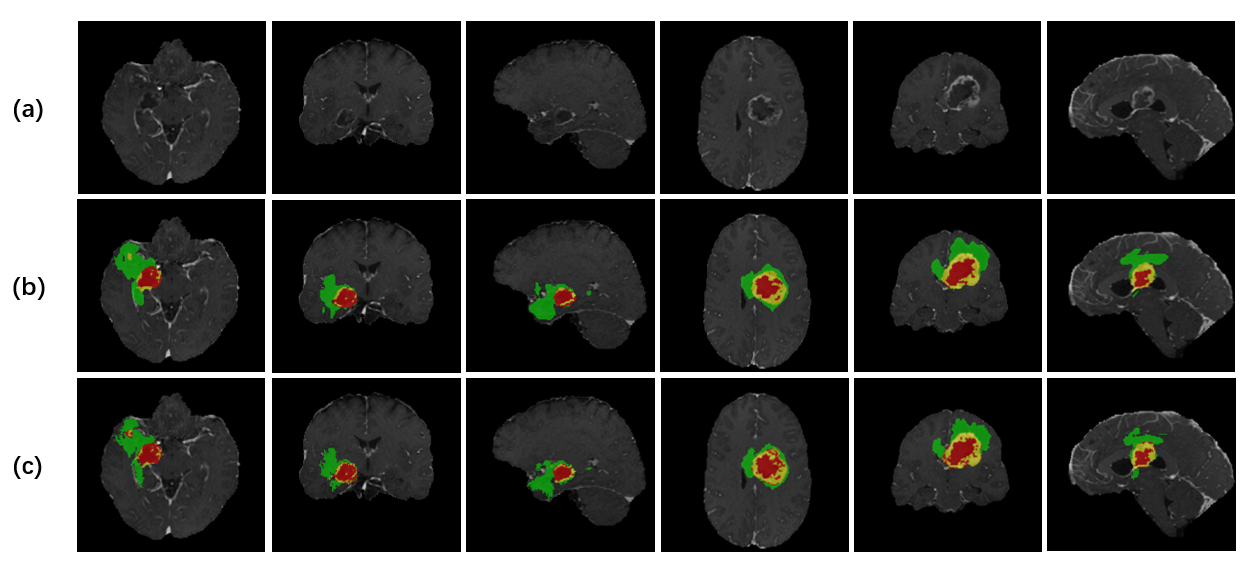}
\caption{Prediction of DMFNet for BraTS 2019 cross-validation on training data. (a) MRI (T1ce), (b) predicted segmentation (c) ground truth.}
\label{fig_train}
\end{figure}

\begin{figure}[ht!]
\centering
\includegraphics[width=125mm]{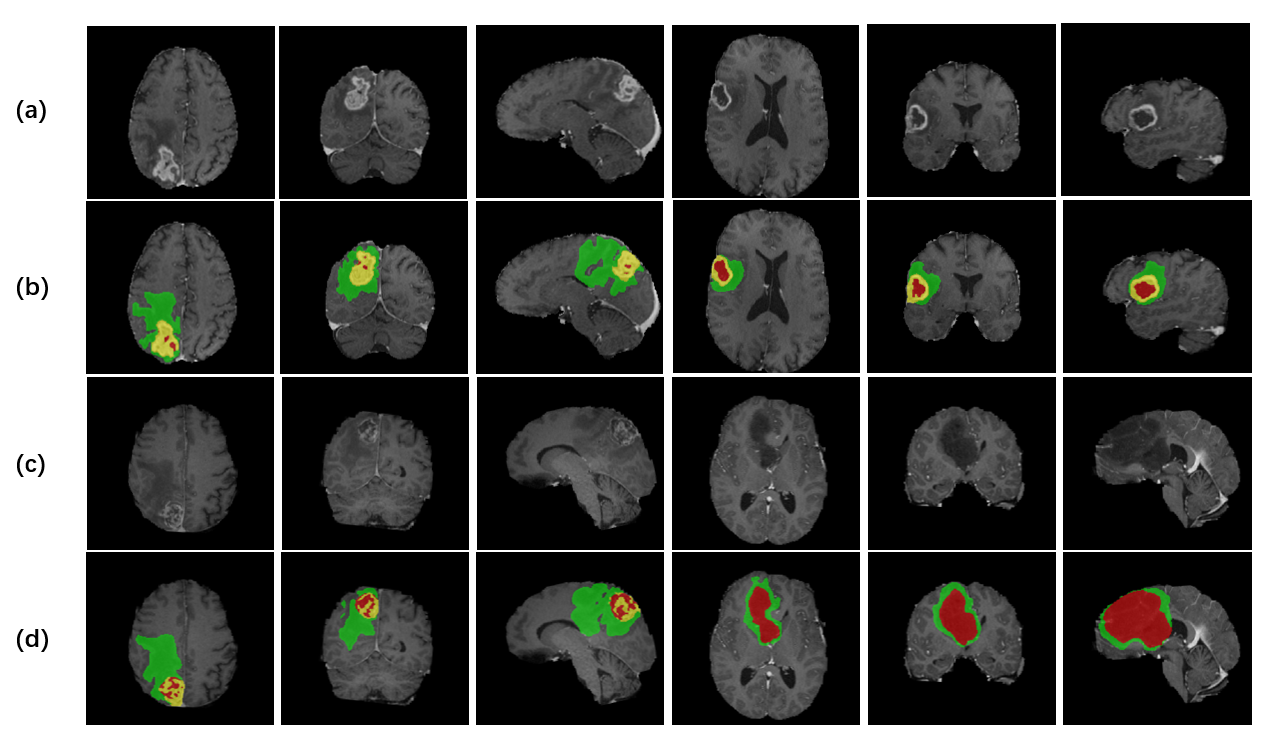}
\caption{Prediction of DMFNet for BraTS 2019 validation and testing data. (a) Validation data; (b) predicted segmentation on (a); (c) testing data; (d) predicted segmentation on (c).}
\label{fig_test}
\end{figure}

\subsubsection{Brain Tumor Segmentation}
For brain tumor segmentation, we first conducted five-fold cross-validation evaluation on the training set, and our DMFNet achieved average dice scores of 80.12\%, 90.62\% and 84.54\% for enhancing tumor (ET), the whole tumor (WT) and the tumor core (TC), respectively. The segmentation results are shown in Fig. \ref{fig_train}, and our results match well with ground truth. Besides, 125 validation cases were evaluated after submitting to the CBICA's Image Processing Portal, achieving average dice scores of 76.88\%, 89.38\% and 81.56\% for ET, WT and TC, respectively. The $3^{rd}~row$ in Table \ref{table:result0} shows the performance metrics that the segmentation network achieved on the testing data. It obtains average dice scores of 78.99\%, 86.71\% and 82.09\% for ET, WT and TC, respectively. We also visualized the segmentation results of DMFNet, as shown in Fig. \ref{fig_test}.

\begin{table}[t]
\caption{Dice and Hausdorff for BraTS 2019 validation and testing dataset.}
\centering
\setlength{\tabcolsep}{0.7mm}{
\begin{tabular}{ccccccc}
\toprule
Dataset & Dice\_ET & Dice\_WT & Dice\_TC & Hausdorff\_ET & Hausdorff\_WT & Hausdorff\_TC\\
\midrule
Validation&76.88\%& 89.38\% & 81.56\%&4.50841&5.03648&6.58191\\
Testing & 78.99\%& 86.71\%& 82.09\%  & 20.24  & 12.45 & 26.99 \\
\bottomrule
\end{tabular}}
\label{table:result0}
\end{table}



\subsubsection{Overall Survival Prediction}

We extracted different features to solve OS prediction problems as follows:
\begin{itemize}
  \item [(a)] 
  Features extracted from PECL-DenseNet   
  \item [(b)]
  handcrafted features
  \item [(c)]
  Position encoding segmentation features
\end{itemize}

Firstly, we trained GBDT regression with 36 handcrafted features. Subsequently, 18 important features were selected by their regression weight and fed into GBDT regression model for training. A comparison of the results obtained from training with 36 and 18 handcrafted features are shown in Table \ref{table:result1}. It is clear that feature selection improves the performance of OS prediction. 

\begin{table}[t]
\caption{36 handcrafted VS 18 handcrafted.}
\centering
\setlength{\tabcolsep}{2.2mm}{
\begin{tabular}{cccccc}
\toprule
Method & Accuracy & MSE & medianSE & stdSE & SpearmanR \\
\midrule
36 handcrafted & 0.31    & 152597.065        & 76777.495   & 192410.851       & -0.091              \\
18 handcrafted & 0.448   & 142485.235        & 64070.727   & 192720.964       & 0.061              \\
\bottomrule
\end{tabular}}
\label{table:result1}
\end{table}

Moreover, we arranged and combined features (a), (b), (c) to train the GBDT regression model, and the corresponding results on validation data were shown in Table \ref{table:result2}. It is obvious that handcrafted features is of great importance compared with the deep features (a) and position encoding features (c) ($2^{nd}~to~4^{th}~rows$). Selecting and combining these three features groups, it is observed that incorporating both handcrafted features and location encoding segmentation data achieved the highest accuracy of 0.586 (Method 6). We then saved the parameters obtained from the regression model that yielded the best result and applied it on testing data (Table \ref{table:result3}). We achieved an accuracy of 0.523 for OS time prediction. 

\begin{table}[t] 
\caption {Validation results of OS prediction with different methods.}
\centering
\scalebox{1.00}{\begin{tabular}{p{1.5cm}<{\centering} | p{0.5cm}<{\centering} p{0.5cm}<{\centering} p{0.5cm}<{\centering}  p{1.6cm}<{\centering} p{1.8cm}<{\centering} c  p{1.8cm}<{\centering} p{1.8cm}<{\centering} p{1.0cm}<{\centering} } 
\toprule 
Method&a&b&c&Accuracy & MSE & medianSE & stdSE & SpearmanR\\ 
\hline
\xrowht{3pt}\\
method 1&${\surd}$&${}$&${}$ &0.379& 431949.975& 270585.384& 488132.042& -0.347\\
\xrowht{6pt}\\

method 2&${}$&${\surd}$&${}$& 0.448& 142485.235& 64070.727& 192720.964& 0.061\\
\xrowht{6pt}\\

method 3&${}$&${}$&${\surd}$& 0.379& 105019.348& 47771.914& 139093.436& 0.07\\
\xrowht{6pt}\\

method 4&${\surd}$&${\surd}$&${}$& 0.483& 118374.49& 68989.292& 132897.288& 0.238\\
\xrowht{6pt}\\

method 5&${\surd}$&${}$&${\surd}$& 0.448& 120356.082& 52497.44& 186701.876& 0.012\\
\xrowht{6pt}\\

method 6&${}$&${\surd}$&${\surd}$& 0.586& 104985.694& 86581.049& 117638.724& 0.218\\
\xrowht{6pt}\\

method 7&${\surd}$&${\surd}$&${\surd}$& 0.517& 200169.575& 51368.509& 309567.261& 0.142\\
\xrowht{3pt}\\
\bottomrule 
\end{tabular}}
\label{table:result2}
\end{table}

\begin{table}[ht]
\caption{Testing results of OS prediction with the method 6.}
\centering
\setlength{\tabcolsep}{4.5mm}{
\begin{tabular}{cccccc}
\toprule
Accuracy & MSE & medianSE & stdSE & SpearmanR \\
\midrule
0.523    & 407196.811        & 55938.713   & 1189657.961       & 0.281    \\
\bottomrule
\end{tabular}}
\label{table:result3}
\end{table}

\section{Conclusion}
In this paper, we utilize DMFNet with weighted dice loss for brain tumor segmentation, which significantly reduces the computation cost and obtains a balance between small and large objects from MRI scans. Segmentations predicted from DMFNet are further utilized to provide explicit tumor information for patient OS prediction. As for OS prediction, GBDT regression is implemented by combining  of deep features derived from the proposed PECL-DenseNet, handcrafted features and position encoding segmentation features. Specifically, we propose a PECL-DenseNet to extract meaningful features, which makes adequate use of unlabeled data and to prevent over-fitting issues. Besides, several clinical features are defined and combined with the deep features from PECL-DEnseNet to train GBDT regression for OS days prediction. Although our methods reveals promising performances for both the brain tumor segmentation and OS prediction tasks, we believe that the performance will be further improved by integrating more MRI modality data and brain tumor molecular information.

%
%
%

\begin{thebibliography}{10}
\providecommand{\url}[1]{\texttt{#1}}
\providecommand{\urlprefix}{URL }
\providecommand{\doi}[1]{https://doi.org/#1}

\bibitem{bakas2017segmentationGBM}
Bakas, S., Akbari, H., Sotiras, A., Bilello, M., Rozycki, M., Kirby, J.,
  Freymann, J., Farahani, K., Davatzikos, C.: Segmentation labels and radiomic
  features for the pre-operative scans of the tcga-gbm collection. the cancer
  imaging archive (2017) (2017)

\bibitem{bakas2017segmentationLGG}
Bakas, S., Akbari, H., Sotiras, A., Bilello, M., Rozycki, M., Kirby, J.,
  Freymann, J., Farahani, K., Davatzikos, C.: Segmentation labels and radiomic
  features for the pre-operative scans of the tcga-lgg collection. The Cancer
  Imaging Archive  \textbf{286} (2017)

\bibitem{bakas2017advancing}
Bakas, S., Akbari, H., Sotiras, A., Bilello, M., Rozycki, M., Kirby, J.S.,
  Freymann, J.B., Farahani, K., Davatzikos, C.: Advancing the cancer genome
  atlas glioma mri collections with expert segmentation labels and radiomic
  features. Scientific data  \textbf{4},  170117 (2017)

\bibitem{bakas2018identifying}
Bakas, S., et~al.: Identifying the best machine learning algorithms for brain
  tumor segmentation, progression assessment, and overall survival prediction
  in the brats challenge. arXiv preprint arXiv:1811.02629  (2018)

\bibitem{von2016search}
von Bartheld, C.S., Bahney, J., Herculano-Houzel, S.: The search for true
  numbers of neurons and glial cells in the human brain: A review of 150 years
  of cell counting. Journal of Comparative Neurology  \textbf{524}(18),
  3865--3895 (2016)

\bibitem{van2010interobserver}
van~den Bent, M.J.: Interobserver variation of the histopathological diagnosis
  in clinical trials on glioma: a clinician’s perspective. Acta Neuropathol.
  \textbf{120}(3),  297--304 (2010)

\bibitem{bray2018global}
Bray, F., Ferlay, J., Soerjomataram, I., Siegel, R.L., Torre, L.A., Jemal, A.:
  Global cancer statistics 2018: Globocan estimates of incidence and mortality
  worldwide for 36 cancers in 185 countries. CA Cancer J Clin  \textbf{68}(6),
  394--424 (2018)

\bibitem{chen20193d}
Chen, C., Liu, X., Ding, M., Zheng, J., Li, J.: 3d dilated multi-fiber network
  for real-time brain tumor segmentation in mri. arXiv preprint
  arXiv:1904.03355  (2019)

\bibitem{feng2018brain}
Feng, X., Tustison, N., Meyer, C.: Brain tumor segmentation using an ensemble
  of 3d u-nets and overall survival prediction using radiomic features. In:
  MICCAI Brainlesion Workshop. pp. 279--288. Springer (2018)

\bibitem{kao2018brain}
Kao, P.Y., Ngo, T., Zhang, A., Chen, J.W., Manjunath, B.: Brain tumor
  segmentation and tractographic feature extraction from structural mr images
  for overall survival prediction. In: MICCAI Brainlesion Workshop. pp.
  128--141. Springer (2018)

\bibitem{lecun2015deep}
LeCun, Y., Bengio, Y., Hinton, G.: Deep learning. Nature  \textbf{521}(7553),
  ~436 (2015)

\bibitem{menze2014multimodal}
Menze, B.H., et~al.: The multimodal brain tumor image segmentation benchmark
  (brats). IEEE Trans. Med. Imag.  \textbf{34}(10),  1993--2024 (2014)

\bibitem{nie2019multi}
Nie, D., Lu, J., Zhang, H., Adeli, E., Wang, J., Yu, Z., Liu, L., Wang, Q., Wu,
  J., Shen, D.: Multi-channel 3d deep feature learning for survival time
  prediction of brain tumor patients using multi-modal neuroimages. Sci. Rep.
  \textbf{9}(1), ~1103 (2019)

\bibitem{perez2017glioblastoma}
P{\'e}rez-Beteta, J., Mart{\'\i}nez-Gonz{\'a}lez, A., Molina, D., Amo-Salas,
  M., Luque, B., Arregui, E., Calvo, M., Borr{\'a}s, J.M., L{\'o}pez, C.,
  Claramonte, M., et~al.: Glioblastoma: does the pre-treatment geometry matter?
  a postcontrast t1 mri-based study. European radiology  \textbf{27}(3),
  1096--1104 (2017)

\bibitem{shattuck2008construction}
Shattuck, D.W., Mirza, M., Adisetiyo, V., Hojatkashani, C., Salamon, G., Narr,
  K.L., Poldrack, R.A., Bilder, R.M., Toga, A.W.: Construction of a 3d
  probabilistic atlas of human cortical structures. Neuroimage  \textbf{39}(3),
   1064--1080 (2008)

\bibitem{wang2015radiological}
Wang, K., Wang, Y., Fan, X., Wang, J., Li, G., Ma, J., Ma, J., Jiang, T., Dai,
  J.: Radiological features combined with idh1 status for predicting the
  survival outcome of glioblastoma patients. Neuro-oncology  \textbf{18}(4),
  589--597 (2015)

\bibitem{wen2016discriminative}
Wen, Y., Zhang, K., Li, Z., Qiao, Y.: A discriminative feature learning
  approach for deep face recognition. In: European conference on computer
  vision. pp. 499--515. Springer (2016)

\bibitem{zhao2018automated}
Zhao, Z., Yang, G., Lin, Y., Pang, H., Wang, M.: Automated glioma detection and
  segmentation using graphical models. PloS one  \textbf{13}(8),  e0200745
  (2018)

\end{thebibliography}
%




\end{document}